\documentclass[
twocolumn,
longbibliography,
showpacs,                     
showkeys,                  
secnumarabic,
amssymb, 
nobibnotes, 
aps, 
pre]{revtex4-1}

\usepackage{graphicx}
\usepackage{latexsym}
\usepackage{amsmath}
\usepackage{verbatim}

\begin{document}

\title{Negative temperatures and the definition of entropy}

\author{Robert H. Swendsen}

\address{Physics Department, Carnegie Mellon University, Pittsburgh, PA 15213}

\author{Jian-Sheng Wang}
\affiliation{Department of Physics,  
National University of Singapore, Singapore 117551, Republic of Singapore}


\date{17 October 2014; revised 12 October 2015}

\begin{abstract}
The concept of 
 negative temperature has recently received renewed interest
 in the context of debates about
the correct definition of 
the thermodynamic  entropy
in  statistical mechanics. 
Several researchers have identified the thermodynamic entropy 
 exclusively with the ``volume entropy''
suggested by Gibbs, and have 
further concluded that by this definition, 
negative temperatures
violate
the principles of thermodynamics. 
We disagree with these conclusions.  
We demonstrate that 
volume entropy
is inconsistent 
with the postulates of thermodynamics 
for systems with 
non-monotonic
 energy densities,
while
a definition of entropy based 
on the probability distributions of
macroscopic variables
does satisfy the postulates of thermodynamics.
Our results confirm 
  that negative temperature is a valid 
  extension of  thermodynamics.
\end{abstract}

\maketitle

\section{Introduction}\label{section: introduction}

In 1951,
Purcell and Pound, and in 1956, 
Ramsey, developed the
principles of negative temperatures 
in thermodynamics and statistical mechanics\cite{Purcell_Pound,Ramsey_Neg_T}.
Their analyses have recently been challenged by 
several authors,
who claim that 
negative temperatures 
are not consistent with thermodynamics and statistical mechanics\cite{Berdichevsky_et_al_1991,
Campisi_SHPMP_2005,
Campisi_Kobe_2010,
Romero-Rochin_2013,Sokolov_2014,DH_Physica_A_2006,DH_2013,DH_NatPhys_2014,DH_reply_to_FW,DH_reply_to_Schneider,HHD_2014,Campisi_PRE_2015,HHD_2015}.
The most clearly stated opposition to negative temperatures    
has been given by Dunkel and Hilbert (DH)\cite{DH_Physica_A_2006,DH_2013,DH_NatPhys_2014,DH_reply_to_FW,DH_reply_to_Schneider}
and Hilbert, H\"anggi, and Dunkel (HHD)\cite{HHD_2014},
so we will primarily address their arguments.
The claims of these authors are based on the assertion 
that  an expression for a
volume  entropy,
which had been suggested by Gibbs\cite{Gibbs},
is the only correct definition of entropy 
in statistical mechanics.

DH and HHD have also claimed that they have extended the 
domain of thermodynamics to very small systems---even systems with a single degree of freedom.
We are very skeptical that any form of thermodynamics 
can apply to small systems,
and we will give our reasons later in this paper.
Our main discussion is restricted to many-body systems,
which is the usual domain of thermodynamics. 
We will not require that the system size be infinite,
but it should be large enough for the relative 
thermal fluctuations
to be smaller than the experimental resolution.  A size of
$10^{12}$ particles,
as might be found in a colloid,
is usually sufficient.

Aside from the issue of small systems, we have two questions to consider:
\begin{enumerate}
\item Are negative temperatures inconsistent with thermodynamics?
\item Is the volume entropy the correct definition of the 
thermodynamic entropy in statistical mechanics?
\end{enumerate}
As we will demonstrate,
the answer to both questions is \textit{no}.

The conclusions
of the opponents of negative temperature
 have already 
been challenged\cite{Landsberg_1959,Rapp_Mandr_Rosch,Braun_et_al,Vilar_Rubi_2014,Frenkel_Warren_2015,Schneider_et_al,Wang_2015,SW_2015_PR_E_R,Buonsante_2015,Poulter_2015,Cerino_2015,Anghel_2015}.
We especially endorse 
the arguments given  
by Frenkel and Warren\cite{Frenkel_Warren_2015},
although
we believe
that  some 
issues still need to be clarified. 
Importantly,
although the discussion has involved 
the question of how to properly define 
the thermodynamic entropy in statistical mechanics,
it has  failed to include  alternative
definitions of entropy
that call for consideration,
as discussed below\cite{RHS_1,RHS_4,RHS_book,RHS_unnormalized}.

The main arguments by the authors 
who 
 oppose negative temperatures and    
advocate the 
volume entropy
concern violations of 
adiabatic invariance 
and related 
claimed inconsistencies with thermodynamics
for a system with a monotonically increasing 
density of states\cite{DH_2013,DH_NatPhys_2014,HHD_2014}.    
In fact,
the violations 
and inconsistencies
they point to are 
all
of order $1/N$,
where $N$ is the number of particles,
and therefore 
disappear into the 
thermal noise,
which is
of order 
$1/\sqrt{N}$.
An important feature of 
macroscopic systems 
is that because thermal fluctuations are so small,
a single measurement of $E$ (energy), $V$ (volume), or $N$ 
will almost certainly produce 
the average value within experimental resolution.
Determining the mean value
with a relative error of less than $1/N$ 
 in the presence of fluctuations would require at least $N$ independent measurements,
 even if each individual measurement were exact.
Even for a colloid with only $10^{12}$ particles,
if an independent measurement could be performed every second,
it would take over 30,000 years to complete the experiment.
The  differences that HHD base their arguments on are unmeasurable.

No arguments have been proposed for the volume entropy 
when the density of states decreases with energy,
which is when negative temperatures might arise.

In Sec.~\ref{section: domain of thermo},
we 
review the postulates of thermodynamics,
and explain the modification
needed to include
negative temperatures.
These postulates are the requirements  that must be satisfied 
by the thermodynamic entropy,
and therefore conditions 
that must be met by
an acceptable definition.  
We discuss the
definitions  considered by DH and HHD
in Sec.~\ref{section: choice of entropies},
and then
 review 
a  definition 
of 
entropy 
based on the thermodynamic postulates 
and the probability of 
the macroscopic variables 
in Sec.~\ref{section: SM to S}.

In Sec.~\ref{section: 1/N},
we show explicitly that 
the inconsistencies
noted  by DH and HHD
are  of order $1/N$,
and  therefore 
negligible for thermodynamic systems.
We also show that there are inconsistencies 
exhibited by the volume entropy 
for systems with non-monotonic densities of states,
and that these inconsistencies 
are of order 1 for systems of all sizes.

In Sec.~\ref{section: critique of HHD},
we demonstrate that the HHD definition of entropy
fails to satisfy the zeroth and second laws of thermodynamics.
In Sec.~\ref{section: Ising model},
we consider 
the interactions of
a system of non-interacting Ising spins, with an inverted energy 
distribution,
with other such spin systems,
or with a system of simple harmonic oscillators.  
This provides an explicit example of a calculation suggested by Frenkel and Warren
for demonstrating that the volume entropy 
violates the second law of thermodynamics\cite{Frenkel_Warren_2015}.
We discuss the applicability of thermodynamics 
to small systems in 
Sec.~\ref{section: small systems}
before summarizing our conclusions 
in the final section.

\section{Thermodynamic conditions for the definition of entropy}
\label{section: domain of thermo}

Thermodynamics was invented in the nineteenth century 
by brilliant scientists 
who did not make use of the concept of atoms.
Most scientists didn't  believe
that molecules existed,
much less that they might provide 
a foundation for the laws of thermodynamics. 
Clausius saw the necessity of entropy
without knowing where it came from 
or what it might mean.
Even after the pioneering work of Boltzmann and Gibbs,
many prominent scientists 
continued to reject 
a molecular explanation 
of macroscopic phenomena.
Although the existence of molecules
is taken for granted today,
it should be remembered that 
the domain of 
thermodynamics 
remains that of macroscopic phenomena
for which individual atoms are not resolved.

Thermodynamics  
ignored fluctuations
from the start:
initially  because the existence of fluctuations 
was not known,
but primarily because
macroscopic states could be completely characterized 
by energy, volume, and particle number.
The now well-known justification for the neglect 
of fluctuations 
is that they are of  order 
$1/\sqrt{N}$,
where $N$ is the number of particles in the system,
and therefore smaller than the resolution of macroscopic measurements.
Even for a small macroscopic system with
$N\sim10^{18}$,
thermodynamic measurements 
rarely have a relative resolution of 
$10^{-9}$.

The requirement that fluctuations are smaller than 
experimental resolution is  fundamental. 
However, it is
not the only requirement.
For two objects to be in thermal contact,
it is necessary for some sort of 
direct interaction 
to exist 
between 
particles in different systems.
Due to the short range of molecular interactions---perhaps a few nanometers---these interacting particles are found only 
on the  interfaces 
where the objects come into contact.
This implies that the 
relative size 
of interactions between three-dimensional objects 
goes roughly as   $N^{-1/3}$.
These surface effects can place a stricter limit on
the domain of validity of thermodynamics,
but one that is still satisfied 
for  macroscopic experiments.

A consequence of these considerations 
is that thermodynamics 
is applicable to finite systems,
as long as 
they  contain  a sufficient number of particles 
for the 
fluctuations
and the 
interface effects
to be neglected.

At least since the work of Callen, 
the structure of
thermodynamics
has been seen to follow logically 
from a small set of postulates,
all of which concern 
 the properties of the thermodynamic entropy\cite{Callen}.
While it won't be necessary to follow the entire development 
of thermodynamics from these postulates,
we will need a subset of them,
which we'll express in a modified form\cite{RHS_book}.

The first postulate is simply the assumption that 
state functions exist
for systems in equilibrium.
A system is regarded 
as being in equilibrium 
if its 
macroscopically 
measurable properties
do not depend on time 
and there is no net transport of 
particles or energy.
State functions
depend 
only on a small number of extensive variables
($E_j$, $V_j$, $N_j$, for the $j$th subsystem).
They do not depend 
on the history of the system.

The second postulate is crucial:
it is a particular  expression of 
the second law of thermodynamics.
It says that there is a state function 
called ``entropy,'' for which:
\begin{quote}
The values assumed by the  extensive  parameters
of an isolated composite system
in the absence of
an internal constraint are those that maximize the  entropy  over
the set of all  constrained macroscopic states\cite{RHS_book}.
\end{quote}
The composite system is,
of course,
assumed to be isolated, so that its total energy, volume,
and particle numbers are constant. 

Since the total entropy of a composite system 
is a maximum after a constraint is released 
and the composite system has come to a new equilibrium,
the change in entropy
during such an irreversible process 
 must be positive.

The third postulate is additivity.
If $S_1(E_1,V_1,N_1)$ and $S_2(E_2,V_2,N_2)$
are the entropies of two systems,
then 
$S_T=S_1+S_2$
is the entropy of the composite system,
whether constraints have been released or not.
This statement ignores any energy 
contributions from  interactions
between particles in different systems.

A consequence of these postulates is that 
if two systems are in thermal contact
(are allowed to exchange heat),
but are  isolated from the rest of the universe,
$E_T=E_1+E_2$
must be constant,
and
we must have 
\begin{equation}\label{dS/dE 0}
\frac{ \partial S_T }{ \partial E_1 }
=
0
=
\frac{ \partial S_1 (E_1)}{ \partial E_1 }
+
\frac{ \partial S_2 (E_T - E_1)}{ \partial E_1 }
\end{equation}
or 
\begin{equation}\label{dS/dE 1}
\frac{ \partial S_1 (E_1)}{ \partial E_1 }
=
\frac{ \partial S_2 (E_2)}{ \partial E_2 }  ,
\end{equation}
where it is implicit that the volumes and the particle numbers are held fixed.
This important result leads directly to 
the zeroth law of thermodynamics.
Since two systems in thermal equilibrium 
with each other have the same temperature,
the derivative 
$\partial S / \partial E$
must be a function of temperature;
this function is usually taken to be $1/T$.
For stability,
the second derivative
$\partial^2 S / \partial E^2$
(holding $V$ and $N$ constant)
must be negative.

The next postulate is usually
the monotonicity of $S$
as a function of $E$,
which allows the function 
$S=S(E,V,N)$
to be inverted to give
$E=E(S,V,N)$.
The key question addressed in this paper
is whether this postulate can 
be abandoned
without running into contradictions.

In the next section
we will review the definitions of entropy 
 considered by 
DH and HHD\cite{DH_Physica_A_2006,DH_2013,DH_NatPhys_2014,HHD_2014,HHD_2015}.
This will be followed in Section \ref{section: SM to S}
by an alternative definition 
that was not considered by HHD\cite{RHS_1,RHS_4,RHS_book,RHS_unnormalized}.
We will argue that  definition satisfies 
all of the postulates of thermodynamics
and should be preferred to 
any of the definitions considered by HHD.

\section{Definitions of entropy used by HHD}
\label{section: choice of entropies}

There are essentially two alternative definitions of the entropy
considered by HHD:

1) The entropy might be defined as the logarithm of a volume in phase space,
which is  known as 
either 
the volume entropy 
or the Gibbs entropy\cite{Gibbs}.
  This is the HHD position.  It has the consequence that
the entropy is given
\textit{in equilibrium}
by
\begin{equation}\label{Gibbs entropy}
S_G = 
k_B \ln
\left[
\frac{1}{h^{3N}}
\frac{1}{N!}
\int \!dp\!  \int \!dq 
\,
\Theta\bigl(E-H(p,q)\bigr)
\right]      ,
\end{equation}
where 
the integrals are over the $3N$ momenta, $p$,
the $3N$ position coordinates, $q$,
$\Theta(\cdot)$
is the step function,
 we have dropped the subsystem index 
for simplicity,
and
Planck's constant $h$
is included to obtain agreement with
quantum results in the classical limit\cite{RHS_book,RHS_unnormalized}.

2) The entropy might be defined as the logarithm of a surface in phase space,
which is known as the surface entropy,
and 
which HHD called the Boltzmann entropy.
We will follow HHD in denoting it 
 by $S_B$. 
This definition gives 
the entropy 
\textit{in equilibrium}
by 
\begin{equation}\label{S B}
S_B(E, V, N) 
=
k_B \ln
\left[
\frac{1}{h^{3N}}
\frac{1}{N!}
\int \!dp\!  \int \!dq 
\,
\delta\bigl(E-H(p,q)\bigr)
\right]      .
\end{equation}

Both of these definitions have the disadvantage that,
due to Liouville's theorem,
 any volume
(or surface)
in phase space 
is time independent 
for systems 
obeying Hamiltonian dynamics\cite{Gibbs}.
This means that 
any function of either a volume or a surface in phase space
will be time independent.
If a constraint in a composite system is released,
the system will undergo an irreversible process
before coming to a new equilibrium state.
The second law requires the entropy to increase,
but 
 both $S_G$ and $S_B$ will remain unchanged
 from their values before  the release of the constraint.

HHD do not address the problem of Liouville's theorem directly,
but instead 
  define the entropy after an irreversible process 
by a different expression,
which requires a new calculation 
in statistical mechanics.
Their new expression is not equivalent 
to the sum of the values of 
$S_G$ for the subsystems,
so it 
 does not obey additivity.
It is also not a function of the energies of individual subsystems,
so that it cannot  predict the equilibrium values of those energies.

In the next section
we will review an alternative  definition of entropy 
in statistical mechanics 
that follows from considering the thermodynamics,
and which obeys the second law
and can be used to predict 
the values of thermodynamic observables
in equilibrium.

\section{From  statistical mechanics to 
the thermodynamic entropy}\label{section: SM to S}

Although  disagreements on 
the existence of 
negative temperatures
center on the definition of entropy
in statistical mechanics,
the predictions of statistical mechanics 
do not rely on 
any  such definition.
Once the probability distribution
in phase space
is given, 
the probability distribution for 
any observables of interest can be computed.
We will use the probability distribution 
for thermodynamic observables 
to justify  a
 definition of entropy
 that is completely consistent with  
 the thermodynamic postulates
 outlined in  Section \ref{section: domain of thermo}\cite{RHS_1,  RHS_4,RHS_book,RHS_unnormalized}.

Consider $M \ge 2$ subsystems
that form an isolated composite system,
with no information about the system's history.
To reduce the proliferation of subscripts we will assume 
that there is only a single type of particle.
Consider any  equilibrium state,
with any combination of 
constraints on the values of the set of  extensive variables
$\{E_j,V_j,N_j \vert\, j=1,2 \dots,M \}$.

If we ignore direct interactions between particles in different subsystems,
the general form of this probability distribution 
is shown in
Ref.~\cite{RHS_unnormalized}
to be proportional to
\begin{equation}\label{W M}
\widehat{W}_M = \prod_{j=1}^M \Omega_j (E_j, V_j, N_j )  ,
\end{equation}
with only the conditions that
$E_{M,\textrm{total}}$,
$V_{M,\textrm{total}}$,
and
$N_{M,\textrm{total}}$
are constant, where
\begin{equation}\label{n conserved}
    E_{M,\textrm{total}} = \sum_{j=1}^M  E_j,  \quad
    V_{M,\textrm{total}} = \sum_{j=1}^M  V_j,   \quad
    N_{M,\textrm{total}} = \sum_{j=1}^M  N_j .  
\end{equation}
The assumption of short-ranged interactions 
is essential in deriving 
Eq.~(\ref{W M}).
If two subsystems are in thermal contact,
the fractional
error due to such interactions 
 is of  order 
 $N_j^{-1/3}$.
 This can be larger than the 
 uncertainty due to fluctuations,
 but still negligible 
 for most macroscopic systems.
 For classical systems in equilibrium,
 we have an explicit expression for the factors in 
 Eq.~(\ref{W M}):
\begin{equation}\label{Omega J}
\Omega_j (E_j, V_j, N_j ) 
=
\frac{1}{h^{3N_j}}
\frac{1}{N_j!}
\int \!dp\!  \int \!dq 
\,
\delta\bigl(E_j-H_j(p,q)\bigr),
\end{equation}
where 
$\int dp$ indicates an integral over all momenta,
and 
$\int dq$ indicates an integral over  configurations
with all particles in the volume 
$V_j$\cite{RHS_1,  RHS_4,RHS_book,RHS_unnormalized}.

Since the probability distributions 
of macroscopic variables---$E_j, V_j, N_j$, 
etc.---are known to have a relative width 
of the order of 
$1/\sqrt{N_j}$
for large $N_j$,
and this is 
assumed to be smaller than experimental error,
it suffices to take the 
equilibrium values 
to be the locations of the maxima.
Since the logarithm is a monotonic function,
$\ln \widehat{W}$
 has its maxima 
at the same locations as 
$\widehat{W}$.
Therefore, 
if we define the entropy of the composite system by
\begin{equation}\label{SM = ln W}
S_M = k_B \ln \widehat{W} + X ,
\end{equation}
it would have its maxima at the equilibrium values of the variables,
and so satisfies the second postulate.
We can also write the total entropy as
\begin{equation}\label{S M}
S_M
=
 \sum_{j=1}^M  S_j (E_j, V_j, N_j) 
 + Y ,
\end{equation}
where  we have included 
additive  constants for generality,
and
\begin{equation}\label{S j}
S_j(E_j, V_j, N_j) 
=
k_B \ln \Omega_j(E_j, V_j, N_j) ,
\end{equation}
or
\begin{equation}\label{S j 2}
S_j(E_j, V_j, N_j) 
=
k_B \ln
\left[
\frac{1}{h^{3N_j}}
\frac{1}{N_j!}
\int \!dp\!  \int \!dq 
\,
\delta\bigl(E_j-H_j(p,q)\bigr)
\right]      .
\end{equation}
Note that this form of the equilibrium entropy
is essentially the same as 
the expression obtained 
from the surface entropy
\textit{in equilibrium},
as given above in 
Eq.~(\ref{S B}).
The most important difference 
for the current discussion is that
defining the entropy from the probability distribution
of the thermodynamic observables
gives the correct increase 
in entropy after an irreversible process\cite{RHS_1,RHS_4,RHS_book,RHS_unnormalized}.
Eq.~(\ref{S j 2})
is also valid for the new equilibrium state
after return to equilibrium.
As discussed in the previous section,
defining entropy 
as a surface or a volume
in phase space 
fails  due to Liouville's theorem\cite{Gibbs}.

An important feature of our definition  is that 
because the entropy is defined as the logarithm of a probability distribution 
for the thermodynamic observables,
the location of the maximum of the entropy 
always corresponds to the mode 
of the probability distribution. 
This means that the predictions of this definition of the entropy
are always correct 
to within the width of the thermal fluctuations.

In the next section
we will compare  
the different equilibrium predictions 
of the  proposed definitions.
Because there is no difference 
\emph{in equilibrium},
between  the predictions of the surface entropy
and our definition 
of  the entropy 
from the probability distribution 
$\widehat{W}_M$
in 
Eq.~(\ref{W M})
we will use the notation $S_B$ 
for both in the following section for simplicity.

\section{Differences in predictions of different entropies}
\label{section: 1/N}

 HHD  have claimed 
that $S_B$
violates equipartition and 
adiabatic invariance
for classical systems of particles
with unbounded energies. 
This is their basis for preferring 
the volume entropy 
for \textit{all} systems,
even those with bounded  energy spectra, 
for which their argument does not apply.
We will analyze their claims
for systems with unbounded energy spectra
in the next subsection,
demonstrating that the 
effect is of order 
$1/N$,
and therefore negligible.
In Subsection \ref{subsection: non-monotonic},
we show that 
$S_B$ again has differences 
between mean and the mode 
of the order of $1/N$,
but that $S_G$
has errors that are of order one
for systems of all sizes.

\subsection{Monotonic energy densities and $1/N$ effects}
\label{subsection: monotonic}

As discussed in 
Sec.~\ref{section: SM to S},
the derivation of
Eq.~(\ref{S j 2}) 
from statistical mechanics
in 
Ref.~\cite{RHS_unnormalized}
assumed that  the equilibrium values of the extensive variables
were given by the location of the maximum 
of the probability distribution (mode).
HHD 
take the position that the correct value 
of an extensive variable in equilibrium
is given by its  mean
over the same probability distribution.
The mean and the mode  generally differ by terms of order
$1/N_j$.
HHD  maintain that 
this difference is an error in $S_B$,
and 
results derived from 
Eq.~(\ref{S j 2}) 
are only an approximation.

Even if the HHD assumption of  equilibrium values 
being 
\emph{exactly equal}
to the mean of the probability distribution
were to be accepted,
the use the mode 
would have to be listed as 
among the best approximations in physics.
As shown in the Introduction,
the difference between the mean and the mode is unmeasurable.


To illustrate these points,
consider an explicit example of 
the origin of the $1/N_j$ differences.
The probability distribution of the energy  
between subsystems $1$ and $2$
 in an ideal gas
 separated by a diathermal wall
  is proportional to\cite{RHS_book}
\begin{equation}\label{CIG energy 1}
E_1^{3N_1/2-1}
\left(E_T - E_1 \right)^{3N_2/2-1},
\end{equation}
where the total energy in the two subsystems is
$E_T=E_1+E_2$.
The \textit{mean} value 
of the energy per particle in subsystem $1$
is  
related to the \emph{mean} energy per particle in subsystem $2$ by
\begin{equation}\label{CIG energy ave 1}
\frac{ \left\langle  E_1  \right\rangle }{ N_1} 
=
\frac{ \left\langle  E_2  \right\rangle }{ N_2}  .
\end{equation}
However,
the mode of the energy,
$E_1^* = E_T - E_2^*$, 
is found by 
setting the derivative of 
Eq.~(\ref{CIG energy 1})
with respect to $E_1$ 
equal 
to zero:
\begin{equation}\label{CIG energy max 1}
\frac{ E_1^* }{ N_1} 
\left( 1 - \frac{2}{3N_1} \right)^{-1}
=
\frac{ E_2^* }{ N_2} 
\left( 1 - \frac{2}{3N_2 }  \right)^{-1}    .
\end{equation}
Thus, HHD are correct
in saying that the mean value of $E_1$ is not 
\textit{exactly}
equal to 
the mode of the probability distribution.
Eq.~(\ref{CIG energy max 1})
exhibits explicitly that this is 
a $1/N_j$ effect.

DH and HHD have argued  against
expressions for the entropy that  
involve an integral over a surface of constant energy in phase space,
 on the grounds that these definitions 
 violate 
adiabatic invariance\cite{DH_Physica_A_2006,DH_2013,DH_NatPhys_2014,DH_reply_to_FW,HHD_2014,HHD_2015}.
Strictly speaking,
 their assertion is
correct, in that such an expression for the  
entropy is not \textit{exactly} constant
during a quasi-static 
adiabatic process.  
However, 
 the violation of adiabatic invariance 
is  also of  order  $1/N_j$,
and therefore negligible 
in comparison with the thermal fluctuations.

The DH and HHD arguments that $S_B$
violates adiabatic invariance 
are also based on 
the energy dependence of the entropy.
The easiest way to see it is to note that 
DH
 have shown that an entropy  
 with an energy dependence of the form 
 $(3N_j /2) k_B \ln E_j$
does satisfy adiabatic invariance,
but one
 with an energy dependence of the form 
 $( 1 - 2/(3N_j))\,(3N_j/2) k_B  \ln E_j  $ 
does not.
The relative difference is obviously proportional to 
$2/(3N_j)$.

For macroscopic systems, 
the
differences of the order $1/N_j$,
on which HHD and DH base their arguments, 
are completely negligible;
indeed, they are unmeasurable,
as shown in the Introduction.

\subsection{Non-monotonic energy densities and large errors for $S_G$}
\label{subsection: non-monotonic}

HHD have disputed the standard thermodynamic result 
that if two systems with equal temperatures 
are brought together,
there will be no net energy transfer.
With the purpose of showing 
 that  
no definition of the entropy always satisfied this condition,
they
examined a non-monotonic density of states
of the form
\begin{equation}\label{HHD density of states}
\omega(E) \propto E (E_m - E)  ,
\end{equation}
where $E_m$ is a constant that specifies the maximum allowed energy.
In  
Fig.~7 of Ref.~\cite{HHD_2014}, 
they plotted the predictions of $S_G$ and $S_B$ for 
the energies of the subsystems at which there would be no net energy transfer.
From the differences in the curves,
they decided that,
although the predictions of $S_G$
were incorrect,
so were the predictions of $S_B$.
They concluded that the prediction of equal partial derivatives
is ``naive''\cite{HHD_2014}.

\begin{figure}[htb]
\includegraphics[width=\columnwidth]{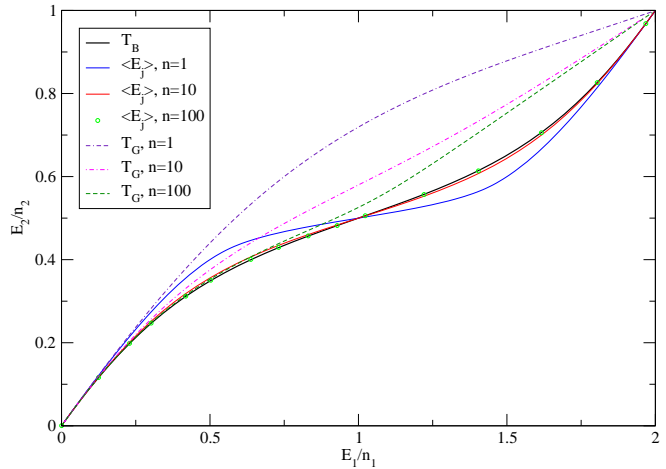}%
\caption{\label{fig:EqualT}
This plot is similar to HHD's figure 7, but with a variable number of degrees of freedom $n$. 
The maximum energies of the two systems were taken to be
$E_1=2n$ and $E_2=n$.
The solid line gives the predictions of $T_B$ for the  mode of the energies of two systems
 such that they will have no net transfer of energy if they are put in thermal contact.
 The other solid curves,
 which are  labelled $\langle E_j \rangle$,
  give the mean of the energies for $n=1, 10,$ and $100$ at zero energy transfer.
It can be seen that for large $n$, the mean and the mode agree very well.
The dashed curves labelled $T_G$ give the energies of the two systems
when they have equal Gibbs temperatures.  
The $T_G$ results show poor agreement with the exact results 
for high energies, that is,
$T_G$ makes  poor predictions of the equilibrium behavior 
when $T_B<0$.
}
\end{figure}

The weakness of their argument is that 
Eq.~(\ref{HHD density of states})
corresponds to a system with only a single degree of freedom.
If a macroscopic system is used for the test,
a more appropriate density of states would be
\begin{equation}\label{HHD density of states n}
\omega(E) \propto E^n (E_m - E)^n  ,
\end{equation}
with a large value of  $n$,
representing a large number of degrees of freedom.
For consistency, the maximum energy should scale with $n$,
that is,  $E_m \propto n$.
In Fig.~\ref{fig:EqualT},
we have redrawn
HHD's Fig.~7 
with a variable number of degrees of freedom.
It can be seen that $T_B$ gives very good predictions
even when $n$ is only as large as $100$.
The small differences go to zero as $1/n$.
In contrast,
the deviation of the predictions of
$S_G$ from the exact curve 
are quite large 
in the high energy region 
for all values of $n$.
As a consequence,
$S_G$ 
fails the HHD test,
while 
$S_B$
calculates the mode exactly  and  
has only an unmeasurably small
deviation from the exact values 
for the mean.

\section{Critique of the HHD interpretation of the laws of thermodynamics}
\label{section: critique of HHD}

HHD claim to offer demonstrations 
that their preferred definition of entropy, $S_G$,
satisfies the laws of thermodynamics.
We disagree, as explained below.

%
\subsection{Volume entropy and the zeroth law}

The zeroth law of thermodynamics
states  that thermodynamic equilibrium is an equivalence relation.
A formulation of the zeroth law due to Planck is that,
``If a body $A$ is in thermal equilibrium with two other bodies 
$B$ and $C$, then $B$ and $C$ are in thermal equilibrium with one another''\cite{Planck_book}.
One reason for the importance  of the zeroth law 
is that it enables us to use a small system  as a thermometer.
In the interpretation of this law,
it should be noted
 that for two systems that are not in thermal contact
to be in equilibrium   with each other means 
that if they were to be brought into thermal contact,
there would be no net flow of energy between them.

In terms of the postulates of thermodynamics,
the zeroth law follows from 
additivity for a composite system and 
the maximization of the total entropy at equilibrium
for systems in thermal contact.
It can be seen immediately by extending 
Eq.~(\ref{dS/dE 1})
to three systems:
\begin{equation}\label{dS/dE 2}
\frac{ \partial S_1 (E_1)}{ \partial E_1 }
=
\frac{ \partial S_2 (E_2)}{ \partial E_2 }  
=
\frac{ \partial S_3 (E_3)}{ \partial E_3 }.  
\end{equation}

HHD make a distinction between what they call
``thermal equilibrium'' and 
``potential thermal equilibrium.''
They restrict thermal equilibrium to systems 
currently in thermal contact, while 
``potential thermal equilibrium''
covers cases in which they have been separated.
The HHD demonstration that $S_G$
satisfies the zeroth law
is limited to systems currently in thermal contact,
and only for systems with unbounded energies.
The reason for this limitation is that
the HHD temperature of a system
is not a property of that system alone;
it is a joint property of the system and 
systems it is in contact with. 
This means that HHD assign a different temperature
$T_G$
 to a system when it is contact with another system, even when the macroscopic properties of the system in question are not changed by such contact.

Consider three systems,
numbered
$1$, $2$, and $3$,
that are in equilibrium with each other 
in the sense given above---that 
there would be no net energy exchange
if they were to be brought into thermal contact with each other.
In Eq.~(33) of Ref.~\cite{HHD_2014}, 
temperatures are defined for each system,
although they do not necessarily have the same values,
as they would in the usual thermodynamics.
Denote these temperatures as $T_1$, $T_2$, and $T_3$.

Now bring systems $1$ and $2$ together.
As shown  in  Eq.~(36)  of Ref.~\cite{HHD_2014},
HHD  define a new temperature for the combined $(1,2)$ system
that depends on  both systems $1$ and~$2$.
The value of this temperature, $T_{1,2}$,
is the same for systems $1$ and $2$
as long as they are in thermal contact with each other,
but it can differ from both $T_1$ and~$T_2$.
In the same way,
HHD define two further temperatures,
$T_{1,3}$ and $T_{2,3}$.
Putting all three systems together gives us yet another temperature,
$T_{1,2,3}$,
for a total of seven values of the temperature
for three systems in thermal  equilibrium
(or ``potential thermal equilibrium'')
 with each other.

Since the HHD temperature is not a state function,
the HHD definition of entropy
is not  state function,
and it
 is not consistent with 
the zeroth law of thermodynamics.

\subsection{First law}\label{subsection: HHD first law}

The first law of thermodynamics is simply the conservation of energy,
which is stipulated to be true 
for all definitions of entropy.
Conservation of energy is automatic with our definition,  which can be readily seen from
Eq.~(\ref{dS/dE 0}).

HHD have claimed that $S_B$ violates the first law in that it violates
the equipartition of energy.
This is the same $1/N$ effect that 
they addressed in connection with adiabatic invariance.
By using the mode instead of the mean for the equilibrium  energy, 
a shift of the order of $1/N$ is introduced.
HHD regard this as important, but we do not.
It is not observable,
being completely obscured by thermal fluctuations.

\subsection{The HHD interpretation of the second law}\label{subsection: HHD 2nd law}

HHD claim that $S_G$ 
satisfies the second law of thermodynamics.
In developing their arguments, 
they consider various formulations of the second law, 
which we will discuss  below. 
 The formulation that they adopt as a test of $S_G$ is, however,
  rather unusual, 
  and does not agree with the second law as it is usually understood.

HHD recognize that $S_G$ 
does not satisfy Clausius'  statement that heat never 
flows spontaneously from a colder to
a hotter body, but they describe 
Clausius' expression as ``naive.''  Indeed,
as we  confirm  in Sec.~\ref{section: numerical test},
$S_G$ does not satisfy this second law in this form,
although the definition of entropy 
in Section \ref{section: SM to S}
   does satisfy it.
The maximum of the sum of the 
volume entropies of two systems 
is not located at the equilibrium values of the energies,
as required by the postulates of thermodynamics.

HHD do accept the Planck formulation
of the second law---that 
the sum of all entropies should not decrease---and they argue 
that this mandate is satisfied by~$S_G$.
We demonstrate that it is not, as follows.

In attempting to demonstrate 
that $S_G$ satisfies Planck's  condition,
HHD consider two systems, $1$ and $2$,
before and after they are brought into thermal contact.
By considering the change in total volume of phase space
used in defining the Gibbs entropies,
they prove that 
if $E_T=E_1+E_2$ is fixed,
\begin{equation}\label{HHD 2nd}
S_{G,1,2} (  E_1 + E_2 ) \ge S_{G,1} (  E_1 ) + S_{G,2} (  E_2 )     .
\end{equation}
A careful examination of their argument shows that this inequality ($\ge$),
which is found in Eq.~(49) of Ref.~\cite{HHD_2014},
should be a strict inequality ($>$).
Note that the maximum of the right-hand side 
as a function of $E_1$ holding $E_T$ fixed
does not lie at the equilibrium values of the energies. 
Beginning with this observation,
consider the following experiment,
which involves systems $1$ and $2$:

\begin{enumerate}
\item 
 Begin with $E_1=E_1^*$ and $E_2=E_2^*$
such that $S_{1,2}^* = S_{G,1} (  E_1^* ) + S_{G,2} (  E_2^* )$
takes on its maximum value,
holding $E_T=E_1+E_2$ constant.

\item 
 Bring systems $1$ and $2$ into thermal contact
with each other,
and let them come to equilibrium.
Denote the new equilibrium energies 
as $E_1^\textrm{eq}$ and $E_2^\textrm{eq}$.
Because the equilibrium values do not coincide with 
the maximum of 
$S_{G,1} (  E_1 ) + S_{G,2} (  E_2 ) $,
this gives the inequality
\begin{equation}\label{HHD 2nd extended prelim}
S_{1,2}^* > S_{G,1} (  E_1^\textrm{eq} ) +
S_{G,2} (  E_2^\textrm{eq} ).
\end{equation}

\item 
We now have the sequence of inequalities:
\begin{equation}\label{HHD 2nd extended}
S_{G,1,2} (  E_1 + E_2 ) \ge S_{1,2}^* > S_{G,1} (  E_1^\textrm{eq} ) +
S_{G,2} (  E_2^\textrm{eq} ).
\end{equation}

\item 
Now separate  systems $1$ and $2$,
so that they are no longer in thermal contact.
They retain the energies $E_1^\textrm{eq}$ and $E_2^\textrm{eq}$.

\item 
If entropy is a state function,
as required by the postulates of thermodynamics,
the total entropy of the system 
has decreased from 
$S_{G,1,2} (  E_1 + E_2 )$ to $S_{G,1} (  E_1^\textrm{eq} ) + S_{G,2} (  E_2^\textrm{eq} )$.
Because of the inequalities in 
Eq.~(\ref{HHD 2nd extended}),
the total Gibbs entropy has decreased,
violating the Planck formulation
of the second law of thermodynamics.

\end{enumerate}

\noindent
This thought experiment demonstrates that $S_G$,
as interpreted by HHD,
violates the second law of thermodynamics.

HHD are aware of this violation,
and they have attempted to evade it with an argument 
in footnote 24\cite{HHD_2014}.
They claim that after separating systems $1$ and $2$
(using our notation),
``two ensembles of subsystems
prepared by such a procedure are not in individual microcanonical
states and their combined entropy remains
$S_{G,1,2} (  E_1^\textrm{eq}  + E_2^\textrm{eq}) = S_{G,1,2}(E_T)$,
so that no violation of the second law has occurred''\cite{HHD_2014}.
They then write that because the energies 
$E_1^\textrm{eq}$ and $ E_2^\textrm{eq}$
are not known exactly,
the reduction of entropy we have pointed to is 
due to ``an additional energy measurement.''

However, their explanation has a serious flaw,
which we can see by 
imagining that the measurement they are making 
does not disturb the system in any way;
the energies are still 
$E_1^\textrm{eq}$ and $ E_2^\textrm{eq}$, \textit{but}
the total entropy has been reduced.
This means that the HHD entropy of system $j$
is not uniquely determined by the values of 
$E_j$, $V_j$, and $N_j$.  In other words, $S_G$,
as interpreted by HHD, is not a state function.
This is a violation of the first postulate of thermodynamics.

In the HHD interpretation of the volume entropy
of two systems in thermal contact is only a function 
of the sum of the energies -- not the energies of the individual systems.
This means that it can make no prediction 
of the equilibrium values of those energies,
so it cannot satisfy the second postulate.

If the volume entropy is interpreted as
remaining the sum of the individual entropies
for systems in thermal contact, 
the location of the maximum of the sum of the entropies of two systems
still does not correspond to the equilibrium values
for systems with non-monotonic energy spectra,
as shown in section \ref{section: 1/N}.

In the next section,
we turn to models with discrete energy levels 
to provide examples 
that  will demonstrate  the consistency of thermodynamic principles with 
descriptions using negative temperatures.
We will also describe
an explicit violation of the second law of thermodynamics
when the volume entropy 
imposes positive temperatures 
on systems with inverted energy distributions.

\section{Models with discrete energy levels}
\label{section: Ising model}

While most of our discussion has concerned general principles,
it is useful to consider models for which explicit calculations 
can be made to demonstrate the relevant ideas.
Models with discrete energy levels 
are particularly well suited for such demonstrations
because calculations and simulations are relatively easy.
In this section, we illustrate the thermal behavior  
of systems with negative temperatures, 
using a 
system of non-interacting Ising spins. 

A system of 
non-interacting spins
 consists of a set of $N$ spins with states denoted
by $\sigma_j$, for $j=1,2, \dots,N$,
where the each spin can take on the values $\sigma_j = +1$ or $\sigma_j=-1$.
The Hamiltonian is 
\begin{equation}
H = - b \sum_{j=1}^N  \sigma_j  ,
\end{equation}
where $b>0$ is a constant proportional to the external magnetic field strength.
The energy of the system at $T=0$ 
(or equivalently, $\beta = 1/ k_B T = \infty$)
is $E=-N b$.
At $T=\infty$ (or $\beta = 0$),
the energy  is $E=0$.
All positive-energy states 
correspond to negative temperatures,
although HD and DHH 
describe them with 
positive values of $T_G$.

The partition function $Z$
of the Ising paramagnet 
can be evaluated explicitly,
and depends only
on the product $\beta b$.
Since the physical properties of the system
are invariant under the transformation
$\beta \rightarrow - \beta$ and
$b \rightarrow - b$,
it is natural to expect the entropy 
to be invariant.

Our first example demonstrates the consistency 
of describing non-monotonic energy densties
with negative temperatures,
as well as the contradictions 
inherent in attempting to impose positive temperatures 
with the Gibbs volume entropy.

\subsection{Two Ising paramagnets with negative temperatures}

We will analyze an experiment with two 
Ising paramagnets,
first using our definition of the entropy 
(and temperature),
and then with the DH and HHD entropy.
Because the sizes of the two systems 
have significant consequences for the 
values of $T_G$, let us assume that $N_1>N_2$.

\subsubsection{Ising paramagnets with negative temperatures}

First consider
 two Ising paramagnets 
in the same magnetic field 
at inverse temperatures
$\beta_1$ and $\beta_2$.
If the two systems are brought into thermal contact,
they will establish a new equilibrium state with a 
common temperature $\beta_f$ that lies between 
$\beta_1$ and $\beta_2$.
This is undisputed for positive values of 
the inverse temperatures,
but the symmetry of the model shows that it 
will also be true 
when the inverse temperatures are negative.
In fact, the new equilibrium temperature, $\beta_f$,
will have the same absolute value in both cases---at 
least when using our definition of entropy.  
Regardless of the definition of entropy,
there is a one-to-one 
correspondence of the probability distributions of the states
in the two cases.

Now suppose that $E_1<0$ ($\beta_1>0$),
but $E_2 >0$  ($\beta_2<0$).
The equations for the probability distributions 
do not change,
and thermal contact between the two systems
will again lead to an equilibrium state with a common 
inverse temperature,
which could be either positive or negative.

The simplicity of this model 
makes it easy to derive the equilibrium 
behavior or carry out a computer simulation.
The average value of any spin in either system
will be given by
$\langle \sigma_j \rangle = \tanh ( \beta b )$, 
where the value of the inverse temperature
$\beta$
is same as what we obtain from our definition of entropy.

\subsubsection{Ising paramagnets described with DH and HHD positive temperatures }

HHD would describe 
Ising paramagnets with positive values of $T_G$,
whether their energies 
were negative or positive.
Let us consider the interesting case of 
$E_1$ and $E_2$ both 
having positive initial values.
If they are brought into thermal equilibrium
with each other and then separated, DH and HHD
have shown that for the case $N_1>N_2$,
$T_{G,1}$ will be greater than $T_{G,2}$ after separation.
This result certainly violates our common understanding of temperature,
and would seem to make the construction of a thermometer impossible.

As discussed in Sec.~\ref{subsection: HHD first law},
HHD have shown that they can construct 
a definition of temperature 
that has the same value for both systems after equilibration,
but only while they remain in contact with each other.

The next subsection 
continues with an example based on a suggestion by Frenkel and Warren
to demonstrate the violation of the second law 
by the Gibbs volume entropy.

\subsection{A numerical demonstration of an argument
due to Frenkel and Warren}
\label{section: numerical test}

Frenkel and Warren suggested an interesting thought experiment 
coupling an Ising model with an ideal gas
to demonstrate that the Gibbs definition of entropy leads 
to a violation of the second law of thermodynamics.\cite{Frenkel_Warren_2015}.
We have taken their idea,
but applied it to interactions between 
a system $1$,
composed of objects with two energy levels,
and a system $2$,
composed of quantum simple harmonic oscillators.
The Hamiltonian of the two-level  system is
\begin{equation}
H_1 = \sum_{i=1}^{N_1} \epsilon\, n_{1,i}, \quad    n_{1,i}= 0, 1,
\end{equation}
where $\epsilon$ is a constant,
and the Hamiltonian for the harmonic oscillator system is
\begin{equation}
H_2 = \sum_{j=1}^{N_2} \hbar \omega \left(   n_{2,j} + \frac{1}{2}\right), \quad  n_{2,j} = 0, 1,2, \ldots.
\end{equation}
For simplicity we take $\epsilon = \hbar \omega$, and we use units in which $\epsilon = k_B=1$.

We have carried out 
 microcanonical Monte Carlo (MC)  computer simulations 
 of the  approach to equilibrium when these two systems
 are brought into thermal contact.
 No assumption concerning the entropy or the temperature  
is necessary for such simulations.
The algorithm simply picks a two-level  object 
and an  oscillator at random.  A proposed move 
changes the energy of the two-level object by $\pm \epsilon$
with equal probability, with a corresponding change
of $\mp \epsilon$ in the energy of the chosen oscillator,
so that the total energy is conserved.  If the proposed move
would take either system to an unphysical state,
the move is rejected.

System $1$ 
was initialized
in its highest-energy state,
that is, $n_{1,i}=1$ for all $i$, with total energy $E_1 = N_1\epsilon = N_1$. 
From the definition of $S_G$ and the HHD finite-difference expression for $T_G$, we have 
\begin{equation}
T_{G,1} 
\approx 
\frac{ \Delta E_1 }{ \Delta S_{G,1}}
= 
\frac{ 1 }{ \ln 2^{N_1} - \ln (2^{{N_1}}-1) }  
\approx
 2^{N_1},
\end{equation}
for sufficiently large $N_1$.

System $2$ was initialized at a value of 
$T_{G,2}$ that was a factor $M$ higher
than $T_{G,1}$. 
Actually,
the initial temperature of system $2$ 
could be taken arbitrarily high 
without affecting our results.

\begin{figure}[htb]
\includegraphics[width=\columnwidth]{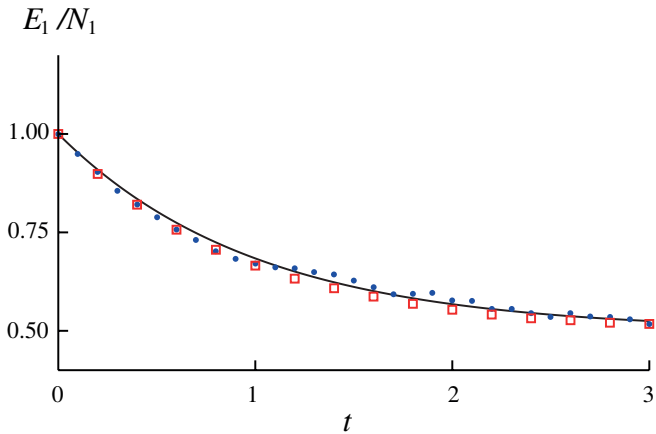}%
\caption{\label{fig:SHO}Energy per particle, $E_1/N_1$, 
of the two-level system vs simulation time~$t$ (in units of MC steps per site).  
Squares: $N_1=5$, $N_2=1$, $M=2$, 
average of $10^4$ runs. Circles: $N_1 = 1000$, $N_2 = 1000$, 
and $T_{G,2} = \infty$, data from a single run.
The solid line is the Markov process prediction, Eq.~(\ref{eqdpdt}).
 }
\end{figure}

We considered both large and small systems,
with the same result in both cases.
For a small system we chose $N_1=5$, $N_2=1$, and $M=2$, so
that the initial Gibbs temperatures are $32$ for the two-level
system and 64 for the system of oscillators.  Figure~\ref{fig:SHO} 
shows the average energy per particle in the two-level system,
which decreases with time despite the fact that the two-level system
has the lower $T_G$.

For a large system we chose $N_1 = 1000$, $N_2 = 1000$,
and  $M=10^6$.  In this case the energy levels of the oscillators 
were too high to store directly in the computer,
but storing these numbers was not necessary. 
Because the total energy of 
the oscillators could not change by more than $N_1$, 
 it was sufficient to store the deviations
$\Delta n_{2,j} $ from the initial values of $n_{2,j}$.
The results of this simulation are also shown in Fig.~\ref{fig:SHO}; 
they are indistinguishable from the results for the small system.

The dynamics of a single two-level particle in this model is very well 
represented by a Markov process, described by the equation
\begin{equation}
\label{eqdpdt}
{ dP_1 \over dt} = \frac{1}{2} P_0 - \frac{1}{2} P_1,\qquad P_0 + P_1 = 1,
\end{equation}
where $P_0$ is the probability that the particle is in the ground state and $P_1$ is the probability that the particle is in the excited state.  
The solution of this equation with the initial condition $P_1(0)=1$, $P_0(0)=0$ gives $P_1(t) = \frac{1}{2}(1 + e^{-t})$, which is also plotted in Fig.~\ref{fig:SHO}.   
This equation is valid as long as the
energies of the oscillators are sufficiently high that no moves are rejected 
for having $n_{2,j}<0$.
This is why our results do not depend on the values used for $M$ or $T_{G,2}$.

As Frenkel and Warren predicted,
these simulations show that when 
the Gibbs entropy and 
$T_G$ are used,
energy can flow from a low-temperature system to a high-temperature system,
increasing the separation of their temperatures\cite{Frenkel_Warren_2015}.
This is a clear violation 
of the Clausius formulation
of the second law.

The next section returns to  the assertion of HHD that
thermodynamics can provide a valid description of small systems---as
small as a single degree of freedom\cite{HHD_2014}.
We show that this claim is not plausible.

\section{Can thermodynamics describe small systems?}\label{section: small systems}

Thermodynamics applies very well to 
 finite systems
 that are large enough 
 to ignore fluctuations and interface effects.
However, 
DH and HHD
have suggested that 
thermodynamics 
should also apply to small systems---even system with only 
a single degree of freedom\cite{DH_Physica_A_2006,DH_2013,DH_NatPhys_2014,DH_reply_to_FW,HHD_2014}.
We are skeptical.

It is certainly true that a thermodynamic point of view  
can be useful in thinking about the behavior of small systems.
Computer simulations have demonstrated clearly that 
small systems can enable us to predict the properties
of much larger systems.
Nevertheless,
the application of thermodynamics to small systems
must be done with care.

There are several finite-size corrections to 
the properties of macroscopic systems.
The most obvious corrections are due to thermal fluctuations,
which are generally of order $1/\sqrt{N}$,
where $N$ is the number of particles in the system.
As mentioned above,
these fluctuations completely mask the
tiny $1/N$ corrections 
that DH and HHD cite as breaking adiabatic invariance.
Even the corrections to Stirling's approximation 
are of order $\ln (N) / N$, which is larger than the 
$1/N$ effects that HHD regard as important.

Normally,  $1/\sqrt{N}$ fluctuations 
are much smaller than the resolution of 
macroscopic measurements
and can be ignored.
To detect 
a $1/N$ effect in the presence of $1/\sqrt{N}$ fluctuations
 would require 
at least $N$ independent measurements, 
each with a resolution better than one part in~$N$.
Even for a colloid with $N \approx 10^{12}$,
this is not feasible.

Whenever two systems come into thermal contact,
there must be a direct interaction 
between the particles in the two systems.
The fraction of such interactions 
is generally of order $N^{-1/3}$, 
which can be larger than the thermal fluctuations 
for small systems.
Certainly,
for a system with only a single degree of freedom,
any coupling to another system 
must correspond to an energy of  roughly of the same magnitude  
as the energy of the system itself.
It cannot be regarded as a small perturbation.

While statistical mechanics can also be applied to the properties of small system,
we can only conclude that thermodynamics 
is a macroscopic theory.

\section{Conclusions}\label{section: summary}

We have shown that the proposal of DH and HHD
that the Gibbs (volume) entropy  ($S_G$)
provides a valid expression for the 
thermodynamic entropy is incorrect.  The volume entropy 
fails to satisfy the postulates of thermodynamics,
the zeroth law of thermodynamics,
and the second law of thermodynamics.  In particular,
we have shown that a description of systems with 
bounded energy spectra and inverted energy distributions
by the volume entropy leads to ambiguous temperatures $T_G$
violating the zeroth law.  Following a suggestion by 
Frenkel and Warren\cite{Frenkel_Warren_2015},
 we have  demonstrated
a spontaneous transfer of energy from a system with lower $T_G$
to a system with higher $T_G$, violating the 
Clausius formulation of the second law.

We have also shown that while the assertions 
of DH and HHD that other expressions for the entropy
violate adiabatic invariance 
are---strictly speaking---correct 
when the density of states is monotonically increasing\cite{DH_Physica_A_2006,DH_2013,DH_NatPhys_2014,DH_reply_to_FW,HHD_2014,HHD_2015},
the effects are of order $1/N$, and are therefore unmeasurable for macroscopic systems.

DH and HHD further claim that thermodynamics can be applied to extremely small
systems, but we have argued that this claim fails to account for fluctuations
and surface interactions.
Thermodynamics is valid for systems in which the 
energy, volume, particle number, etc.\ 
have sufficiently small fluctuations
that their average values 
are enough to fully characterize the (macroscopic)  
state of the system for the experiments of interest.

We have provided support for 
the  proposal 
by Purcell and Pound and by Ramsey    
for the use of 
negative temperatures 
to describe systems with bounded energies
and non-monotonic energy densities.

\section*{Acknowledgements}
We would like to thank Roberta Klatzky for helpful discussions.
We would also like to thank 
Daniel V. Schroeder for several useful comments.

\makeatletter
\renewcommand\@biblabel[1]{#1. }
\makeatother

\bibliography{Entropy_citations_3}

\end{document}